# Temperature-Induced Changes in Optical Properties of Cubic $CH_3NH_3SnI_3$ Using Phonopy-FHI-aims

Hassan Abdulsalam[1*], Fatima Musa Lariski[1], Habiba Garba Ahmad[2] and Hassana Muhammad Shuwa[1]

[1.]Department of Physics,
Yobe State University,
P.M.B.1144
Damaturu, Yobe
Nigeria.

[2]Department of Physics,
Federal College of Education,
Panshekara Rd, Kofar Kabuga,
Kano
Nigeria.

Email: ahassanabdulsalam@gmail.com

## Abstract

*This research explores how temperature influences the optical characteristics of Cubic Methylammonium Tin Iodide ($CH_3NH_3SnI_3$), a material showing great promise for solar cell technology. The variations in the optical absorption coefficient, refractive index, extinction coefficient, reflectivity, and optical conductivity was examined over a range of temperatures and crystallographic orientations ([100], [010], [001]). The findings indicate that as temperature rises, there is a general decrease observed in the absorption coefficient, refractive index, extinction coefficient, and optical conductivity. Similarly, reflectivity also tends to decrease with increasing temperature. These findings suggest that Cubic $CH_3NH_3SnI_3$ exhibits consistent transparency, a stable refractive index, and relatively high reflectivity across different temperature conditions. Its low optical conductivity, typical of semiconductor materials, indicates its suitability for facilitating efficient charge separation and transport within solar cells. This research significantly adds to our comprehension of the optical properties of $CH_3NH_3SnI_3$, paving the way for its potential use in solar cell applications.*

## INTRODUCTION

Perovskite materials, including methylammonium lead halides, offer cost-effective production and straightforward manufacturing processes. Solar cell efficiencies of devices using these materials have increased from 3.8% in 2009(Kojima et al., 2009) to a certified 20.1% in 2014, making this the fastest-advancing solar technology(Völker et al., 2015). According to detailed balance analysis, the efficiency limit of perovskite solar cells is about 31%, which approaches the Shockley-Queisser of gallium arsenide which is 33% (Sha et al., 2015).Their

---

high efficiencies and low production costs make perovskite solar cells a commercially attractive option. At approximately 330 K, $CH_3NH_3PbI_3$ adopts a cubic crystal structure. As the temperature drops to around 236 K, the cubic phase undergoes a transformation into the tetragonal phase. As the temperature decreases lower to about 177 K, the tetragonal phase is transformed into orthorhombic crystal systems(Oku, 2015). The relationship between temperature and the crystal structures of $CH_3NH_3PbI_3$ demonstrates the significant impact of temperature on the electronic properties, including energy band gaps, of perovskite-type compounds. While methylammonium lead iodide ($CH_3NH_3PbI_3$) has shown remarkable efficacy as a photovoltaic material, concerns persist regarding the toxicity of lead. Lead-based perovskites are a major issue that may prejudice implementation of any PSC technology, both regulation and common sense suggest that PSCs have to become lead free to deliver a sustainable technology(Abate, 2017). The pursuit of lead-free halide perovskites is of paramount importance. $Sn^{2+}$ metal cations emerge as a prominent alternative to $Pb^{2+}$ within the perovskite structure due to their analogous $s^2$ valence electronic configuration to $Pb^{2+}$. $Sn^{2+}$ can form a perovskite with a basic formula $ASnX_3$ (A= $CH_3NH_3$ and X = halide) because the ionic radius of $Sn^{2+}$ is similar to that of $Pb^{2+}$(Abate, 2017). Given the noted similarity, methylammonium tin iodide ($CH_3NH_3SnI_3$) stands out as a frequently considered substitute for $CH_3NH_3PbI_3$ in the production of solar cells and optoelectronic devices. However, optical properties are sensitive to the lattice structure and dynamics, which can be affected by external factors such as temperature, pressure and strain(Zhang et al., 2018). Sabetvand and co explore the impact of temperature, specifically at 275 K, 300 K, and 325 K, on the electronic and optical characteristics of $CH_3NH_3SnI_3$ perovskite. Employing a blend of molecular dynamics and density functional theory techniques, they contrasted the properties of the standard (300 K) structure with those of the expanded (325 K) and contracted (275 K) structures. The results suggest that $CH_3NH_3SnI_3$ has potential for optoelectronic applications and better optical performance at higher temperatures as an absorber layer in solar cells(Sabetvand et al., 2020). In this work, the effect of temperature [for range of $-20\ °C\ to\ 50\ °C$ in a step of 10 °C] on the optical properties of cubic $CH_3NH_3SnI_3$ with respect to [100], [010] and [001] orientations were investigated using phonopy-FHI-aims code.

In an ideal harmonic system, which is fully determined by the dynamical matrix $\boldsymbol{D(q)}$, its Hamiltonian does not depend on the volume, this implies that the harmonic Hamiltonian is independent of the lattice parameters, and as a consequence of this, the lattice expansion coefficient $\alpha(T)$ vanishes (Ashcroft and Mermin, 1976).

$$\alpha(T) = \frac{1}{a}\left(\frac{\partial a}{\partial T}\right)_P \qquad (1)$$

To ascertain the temperature dependency of the optical properties, it is essential to assess the lattice expansion. The quasi-harmonic approximation is used to account for the anharmonic effects in the determination of the lattice expansion(Biernacki and Scheffler, 1989). The usage of the quasi-harmonic approximation requires the determination of how the phonons, i.e., the vibrational band structures and the associated free energies, change with the volume(Christian, 2015).

**METHODOLOGY**
In this work ab initio calculations with density functional theory (DFT), using the FHI-aims software package was carried out (Blum et al., 2009). The study utilized the BLYP parameterization of the Generalized Gradient Approximation (GGA) to assess the exchange-correlation energy. The investigation focused on determining how variations in the volume of materials impact their vibrational band structures and related free energies. The optimal lattice constant for the materials was determined through the use of the Birch-Murnaghan equation of state, which involved identifying the minimum values of both the total energy $E_{DFT}(V)$ and free energy $F^{ha}(T,V)$(Franz and Sergey, 2013). In the canonical ensemble, the relevant thermodynamic potential that needs to be minimized is the free energy $F(T,V)$ which is given by:

$$F(T,V) = E_{DFT}(V) + F^{ha}(T,V) \qquad (2)$$

To account for the volume's impact on $F(T,V)$,), the total energy $E_{DFT}(V)$, and the Helmholtz free energy, $F^{ha}(T,V)$, were computed across a range of lattice constants. Subsequently, the Birch-Murnaghan Equation of State was employed using the Phonopy program package and its FHI-aims interface to evaluate and minimize equation (2). To carry out the procedure described above, a FHI-aims python script called *Compute_ZPE_and_lattice_expansion.py* (Christian, 2015) was employed, which requires inputs such as the optimal lattice constant, temperature range, and geometry information. The script generates an output file containing temperature, lattice constant, and lattice expansion coefficient.

The frequency-dependent complex dielectric function ε(ω) plays a pivotal role in understanding the energy band structure of solids (Dresselhaus, 2001). From complex dielectric function $\varepsilon(\omega)$, various optical properties can be derived, including the refractive index (n(ω)), extinction coefficient (k(ω)), absorption coefficient (α(ω)), reflectivity (R(ω)), and optical conductivity (σ(ω)) (Saha et al., 2000).

i. The absorption coefficient, α(ω), quantifies the extent to which light of a particular wavelength can penetrate a material before being absorbed. Materials with low absorption coefficients appear transparent to that wavelength, particularly if they are thin. The absorption coefficient is determined by both the material itself and the wavelength of the incident light (Wang, 2008).

$$\alpha(\omega) = \frac{\sqrt{2}\,\omega}{c}\sqrt{\left(\sqrt{\left(\varepsilon_1(\omega)\right)^2 + \left(\varepsilon_2(\omega)\right)^2}\right) - \varepsilon_1(\omega)} \qquad (3)$$

ii. The optical conductivity, σ(ω), relates the current density to the electric field across a material for various frequencies. It's a broader concept than electrical conductivity, encompassing a wider frequency range (Gervais, 2002).

$$\sigma(\omega) = \frac{\omega}{4\pi}\varepsilon_2(\omega) \qquad (4)$$

iii. The extinction coefficient, k(ω), measures the rate of decrease in transmitted light due to scattering and absorption in a medium. It quantifies the attenuation of an electromagnetic wave as it traverses through the medium (Valizade et al., 2019).

$$k(\omega) = \sqrt{\frac{\sqrt{\left(\varepsilon_1(\omega)\right)^2 + \left(\varepsilon_2(\omega)\right)^2} - \varepsilon_1(\omega)}{2}} \qquad (5)$$

iv. Reflectivity, R(ω), denotes the fraction of incident light that is reflected from a material's surface relative to the incident light intensity. It characterizes the material's ability to reflect light both at the surface and within its volume (Ujihara, 1972).

$$R(\omega) = \frac{(n(\omega)-1)^2 + k(\omega)^2}{(n(\omega)+1)^2 + k(\omega)^2} \qquad (6)$$

v. The refractive index, n(ω), quantifies the degree of light bending as it transitions between different media. It's a measure of the ratio of light velocity in a vacuum to its velocity in a substance (Afromowitz, 1974), indicating the medium's optical density.

$$n(\omega) = \sqrt{\frac{\sqrt{(\varepsilon_1(\omega))^2 + (\varepsilon_2(\omega))^2} + \varepsilon_1(\omega)}{2}} \qquad (7)$$

To compute the frequency-dependent complex dielectric function in FHI-aims, the "compute_dielectric" tag was added to the control.in file. This produced output files containing the real and imaginary components ($\varepsilon_1$ and $\varepsilon_2$) of the frequency-dependent complex dielectric function. For investigating the temperature dependence of the frequency-dependent complex dielectric function, a second Python script was utilized. This script calculated the frequency-dependent complex dielectric function for geometries constructed using the output (which includes temperature, lattice constant, and lattice expansion) of the first script. The calculations were performed along the [100], [010], and [001] cubic directions for a temperature range from -20°C to 50°C in increments of 10°C. The absorption coefficient α(ω), optical conductivity σ(ω), extinction coefficient k(ω), reflectivity R(ω), and refractive index n(ω) were then obtained from the frequency-dependent complex dielectric function using equations 3 to 7, respectively.

**Results and Discussion**

*Optical Absorption Coefficient*
The optical absorption coefficient of a material provides crucial insights into its potential for solar energy conversion efficiency, making it a vital parameter for assessing a material's suitability for practical application in solar cells. The calculated optical absorption spectra are depicted in Figure 1.

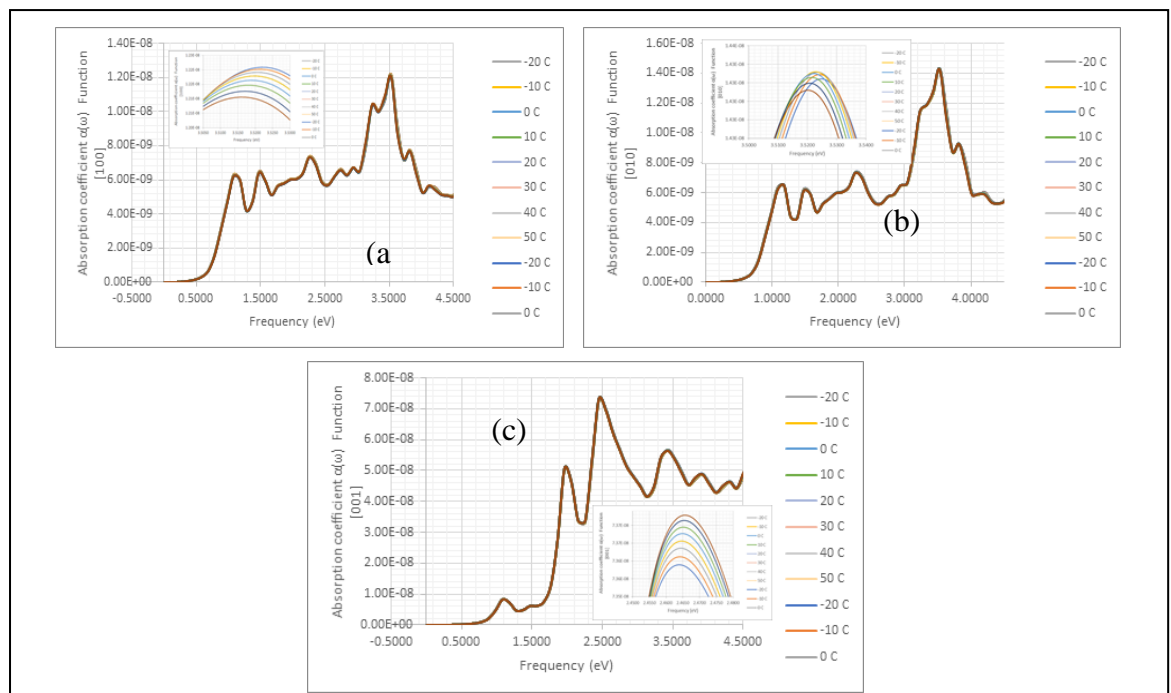

**Figure 1** Graph of Absorption coefficient α(ω) function against Frequency for (a) [100] (b) [010] and (c) [001]

Table 1 displays the absorption coefficient spectrum's highest peaks at each temperature for the crystallographic orientations, derived from the plots in Figure 1. It is observed that there is decrease in the absorption coefficient with increasing temperature (-20°C to 50°C) across all crystallographic orientations. This behaviour suggests a reduction in the material's absorptive capacity with temperature rise. The absorption coefficients are quite small, indicating that the material is not highly absorptive to the incident electromagnetic radiation (Honsberg and Bowden, 2019). Comparing this with the literature, the study on cubic $CH_3NH_3SnI_3$ perovskite by Sabetvand et al. (2020) indicates a similar trend, with enhanced optical performance at higher temperatures. However, the study on Formamidinum based perovskite (Wang et al., 2020) suggests that $FAPbI_3$ exhibits higher absorption compared to Methylammonium based perovskites.

**Table 1 Temperature and Absorption coefficient spectrum (Highest Peak)**

| *Temperature* (°C) | *Absorption coefficient spectrum* (*Highest Peak*) | | |
|---|---|---|---|
| | [100] | [010] | [001] |
| -20 | 1.218E-08 | 1.430E-08 | 7.318E-08 |
| -10 | 1.217E-08 | 1.431E-08 | 7.320E-08 |
| 0 | 1.216E-08 | 1.431E-08 | 7.321E-08 |
| 10 | 1.214E-08 | 1.430E-08 | 7.322E-08 |
| 20 | 1.212E-08 | 1.430E-08 | 7.323E-08 |
| 30 | 1.209E-08 | 1.429E-08 | 7.324E-08 |
| 40 | 1.206E-08 | 1.428E-08 | 7.325E-08 |
| 50 | 1.203E-08 | 1.426E-08 | 7.325E-08 |

*Refractive Index*

The knowledge of refractive index is important for the applications of a material in optical devices like solar cell, photo-detector, waveguide and photonic crystal(Hoffman et al., 2016). Figure 2 contains the computed refractive index spectra, while Table 2 presents the static refractive index and temperature for the crystallographic orientations obtained from Figure 2.

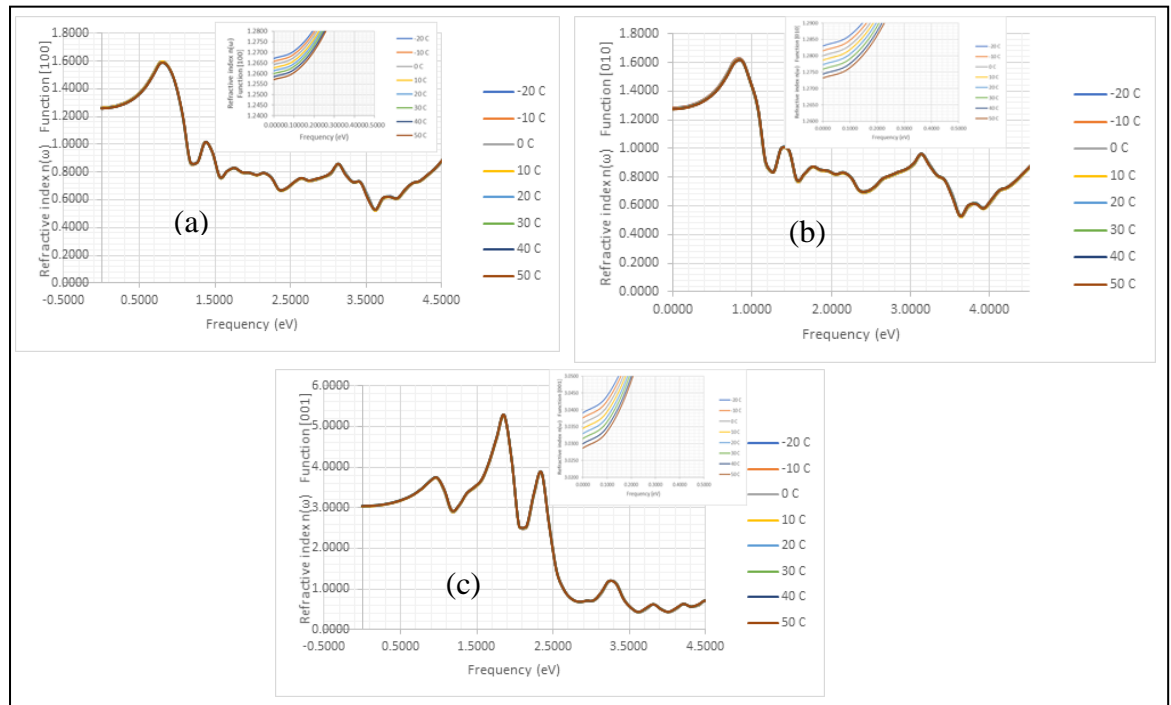

**Figure 2 Graph of Refractive index n(ω) function against Frequency for (a) [100] (b) [010] and (c) [001]**

With the temperature ascending from -20°C to 50°C, the static refractive index diminishes for all three orientations [100], [010], and [001]. Notably, the values of the refractive index exhibit a smooth variation in response to temperature changes. The refractive index changes with temperature, which is a typical behavaior for many materials (Tan, 2021). This temperature dependence is crucial in various applications, such as optical devices and sensors. The refractive index values generally fall within the range of approximately 1.257 to 1.267 for the [100] orientation, 1.273 to 1.283 for the [010] orientation, and 3.029 to 3.039 for the [001] orientation, Similar to Lithium Niobate, $CH_3NH_3SnI_3$ exhibits anisotropic optical properties, meaning the refractive index varies with the crystal orientation (Kogelnik and Shank, 1971). Comparing these results with existing literature, the study by Sabetvand et al. (2020) on $CH_3NH_3SnI_3$ perovskites indicates a decrease in refractive index at higher temperatures, which is in line with the behaviour observed in the material studied here.

Table 2 Temperature and Static Refractive index $n(\omega = 0)$

| Temperature (°C) | Static Refractive index $n(\omega = 0)$ | | |
|---|---|---|---|
| | [100] | [010] | [001] |
| -20 | 1.267 | 1.283 | 3.039 |
| -10 | 1.266 | 1.282 | 3.038 |
| 0 | 1.264 | 1.280 | 3.036 |
| 10 | 1.263 | 1.279 | 3.034 |
| 20 | 1.261 | 1.277 | 3.033 |
| 30 | 1.260 | 1.276 | 3.031 |
| 40 | 1.258 | 1.274 | 3.030 |
| 50 | 1.257 | 1.273 | 3.029 |

*Extinction index*

The extinction coefficient values reflect the intensity of absorption, with the extinction index spectra depicted in Figure 3. Table 3 displays the highest peaks of the extinction index spectra at each temperature for the respective crystallographic orientations, as obtained from Figure 3. The extinction coefficient generally decreases for all three orientations [100], [010], and [001], as the temperature increases, this means that as the temperature rises, the material tends to absorb less light or electromagnetic radiation (Tip, 2008). As the temperature increases, the value of the extinction coefficient decreases, eventually reaching a relatively constant value (Sabetvand et al., 2020).

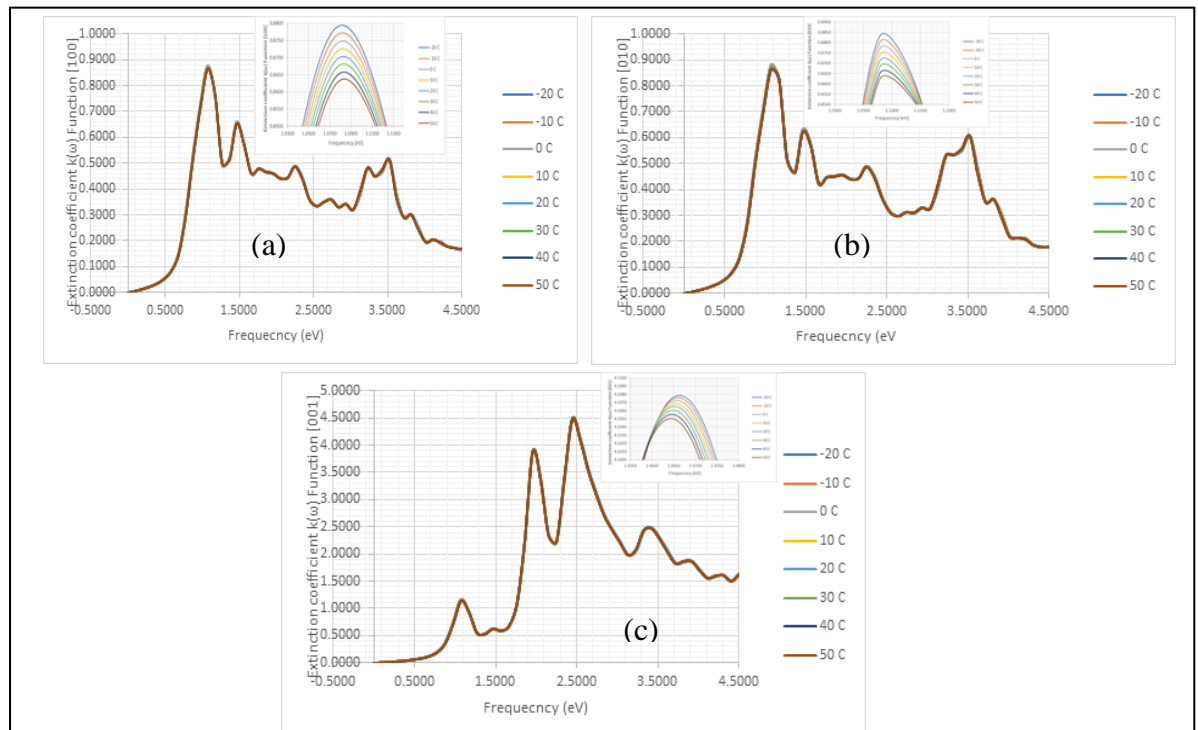

**Figure 3** Graph of Extinction coefficient k(ω) function against Frequency for (a) [100] (b) [010] and (c) [001]

At the highest peak, the extinction coefficient values are relatively large, especially for the [001] orientation. The values are in the range of approximately 0.863 to 0.879 for the [100] orientation, 0.862 to 0.883 for the [010] orientation, and 4.479 to 4.483 for the [001] orientation, this suggests that the extinction coefficient values vary depending on the crystal orientation (Sabetvand et al., 2021).

**Table 3 Temperature and Extinction coefficient spectrum (Highest Peak)**

| *Temperature* (°C) | *Extinction coefficient spectrum* (*Highest Peak*) | | |
|---|---|---|---|
| | [100] | [010] | [001] |
| -20 | 8.793E-01 | 8.829E-01 | 4.479E+00 |
| -10 | 8.770E-01 | 8.800E-01 | 4.480E+00 |
| 0 | 8.745E-01 | 8.768E-01 | 4.480E+00 |
| 10 | 8.722E-01 | 8.738E-01 | 4.481E+00 |
| 20 | 8.699E-01 | 8.708E-01 | 4.482E+00 |
| 30 | 8.676E-01 | 8.679E-01 | 4.482E+00 |
| 40 | 8.652E-01 | 8.648E-01 | 4.483E+00 |
| 50 | 8.631E-01 | 8.621E-01 | 4.483E+00 |

*Reflectivity*

Reflectivity stands as another critical optical property relevant to solar cells, representing the portion of incident light that undergoes reflection. Figure 4 showcases the computed refractive index spectra, while Table 4 delineates the static reflectivity alongside corresponding temperatures for the crystallographic orientations, as derived from Figure 4.

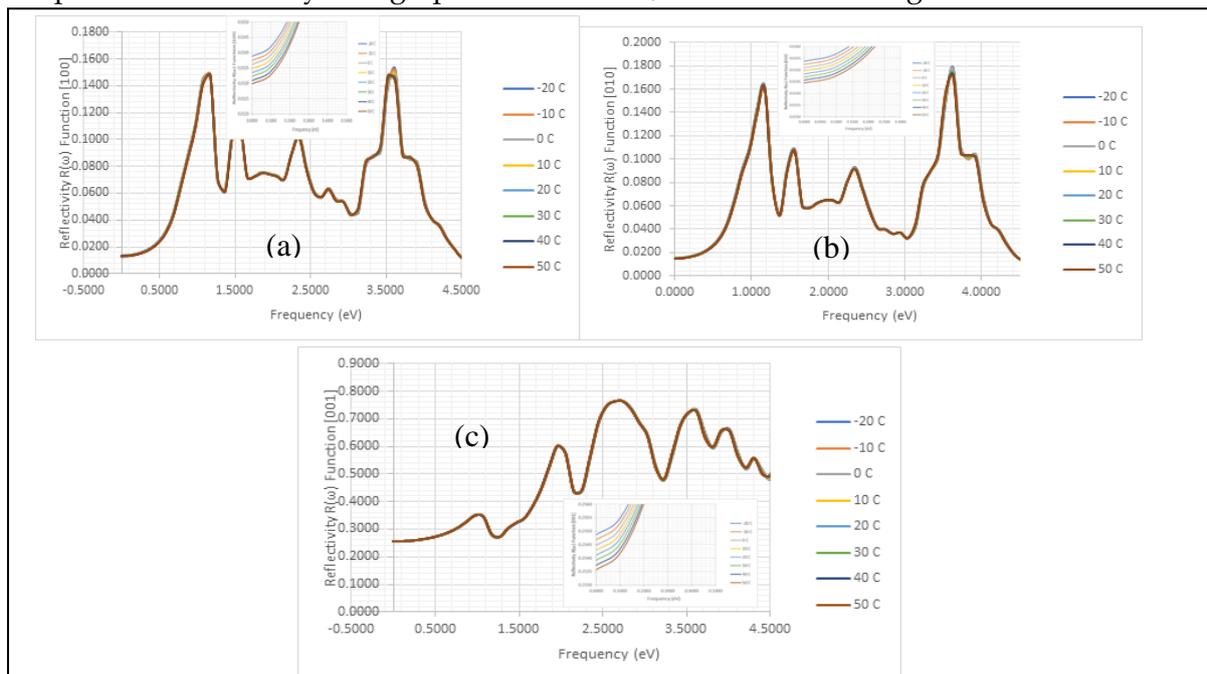

**Figure 4 Graph of Reflectivity R(ω) function against Frequency for (a) [100] (b) [010] and (c) [001]**

From -20°C to 50°C, the static reflectivity generally decreases for all three orientations [100], [010], and [001]. This means that as the temperature rises, the material tends to reflect less incident light (Sabetvand et al., 2020). At $\omega = 0$, the static reflectivity values are relatively small, and they are in the range of approximately 0.01298 to 0.01388 for the [100] orientation, 0.01445 to 0.01537 for the [010] orientation, and 0.2536 to 0.2549 for the [001] orientation, these reflectivity values provide insights into how the material interacts with light at zero frequency, and the variations among different orientations highlight its anisotropic behaviour (Olmon et al., 2012). In essence, the intricate interplay between temperature and reflectivity in absorbing Bragg reflectors provides valuable insights into light-matter interactions. It underscores the importance of considering temperature effects in the design and implementation of optical devices for various applications (Shen et al., 2001) (Tsuda et al., 1996).

**Table 4 Temperature and Static Reflectivity $R(\omega = 0)$**

| *Temperature* (°C) | *Static Reflectivity $R(\omega = 0)$* | | |
|---|---|---|---|
| | [100] | [010] | [001] |
| -20 | 1.388E-02 | 1.537E-02 | 2.549E-01 |
| -10 | 1.375E-02 | 1.523E-02 | 2.547E-01 |
| 0 | 1.361E-02 | 1.509E-02 | 2.545E-01 |
| 10 | 1.348E-02 | 1.496E-02 | 2.543E-01 |
| 20 | 1.335E-02 | 1.482E-02 | 2.541E-01 |
| 30 | 1.322E-02 | 1.470E-02 | 2.539E-01 |
| 40 | 1.309E-02 | 1.456E-02 | 2.537E-01 |
| 50 | 1.298E-02 | 1.445E-02 | 2.536E-01 |

*Optical Conductivity*

The optical conductivity spectrum represents the material's response to incident light at different frequencies. the conductivity spectrum spectra were illustrated in Figure 5, while Table 5 presents the conductivity spectrum spectra highest peaks at each temperature for the crystallographic orientations obtained from Figure 5.

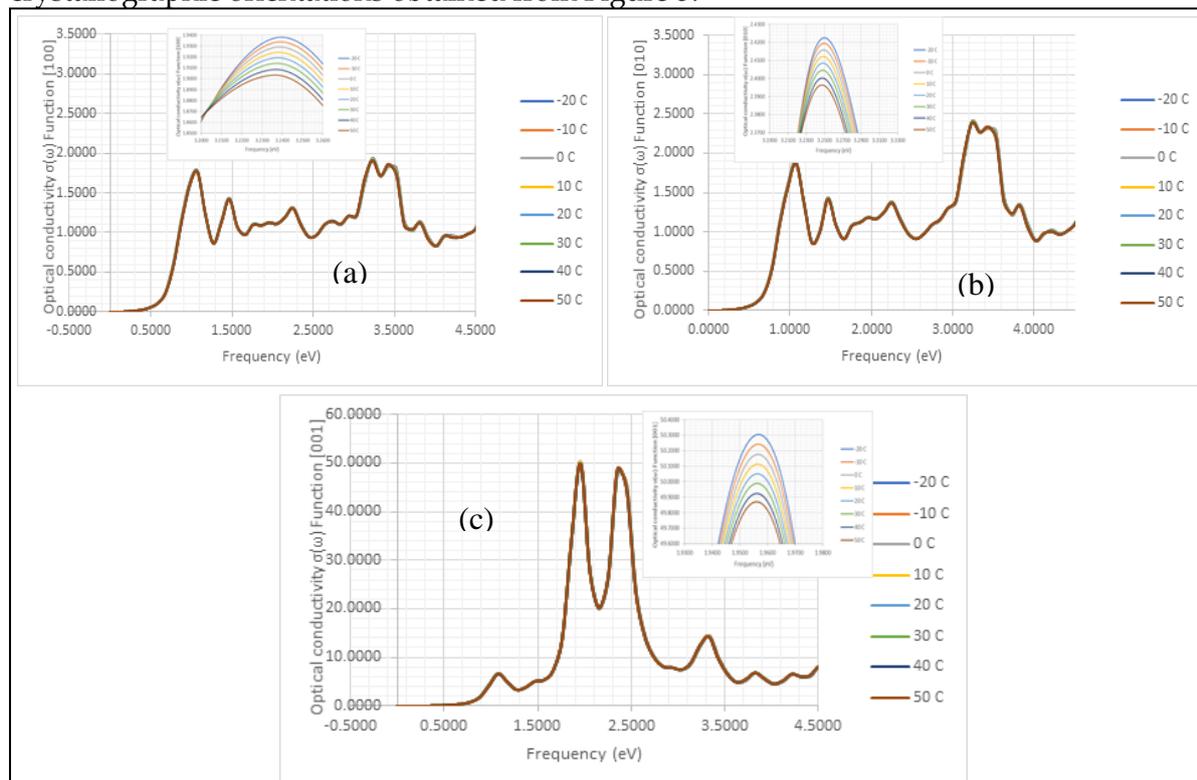

**Figure 5** Graph of Optical conductivity σ(ω) function against Frequency for (a) [100] (b) [010] and (c) [001]

As the temperature rises from -20°C to 50°C, the optical conductivity generally decreases for all three orientations [100], [010], and [001]. This indicates that as the temperature increases, the material tends to conduct less electric current in response to incident light(Sabetvand et al., 2020). Notably, at the highest peak in the spectrum, the values are relatively high, particularly for the [001] orientation, where the optical conductivity reaches its peak values. It implies that this orientation is particularly conducive to converting light energy into electrical current, enhancing the overall performance and efficiency of devices utilizing this material(Takahashi et al., 2011, Iefanova et al., 2016).

**Table 5 Temperature and Optical Conductivity Spectrum (Highest Peak)**

| *Temperature* (°C) | *Optical conductivity spectrum* (*Highest Peak*) | | |
|---|---|---|---|
| | [100] | [010] | [001] |
| -20 | 1.937 | 2.409 | 50.243 |
| -10 | 1.933 | 2.407 | 50.178 |
| 0 | 1.928 | 2.405 | 50.111 |
| 10 | 1.924 | 2.402 | 50.044 |
| 20 | 1.919 | 2.399 | 49.980 |
| 30 | 1.914 | 2.396 | 49.915 |
| 40 | 1.908 | 2.392 | 49.847 |
| 50 | 1.903 | 2.389 | 49.789 |

**CONCLUSION**

This study delved into the influence of temperature on the optical properties of Cubic Methylammonium Tin Iodide ($CH_3NH_3SnI_3$), encompassing the optical absorption coefficient, refractive index, extinction index, reflectivity, and optical conductivity. These properties were meticulously analysed across a spectrum of temperatures and crystallographic orientations ([100], [010], [001]). The findings revealed a consistent decrease in the absorption coefficient with increasing temperature across all orientations, indicating a relatively uniform absorption behaviour. Moreover, the static refractive index exhibited a gradual decrease as temperature rose, a trend commonly observed in various materials. Extinction coefficients followed suit, declining with temperature increments, suggesting diminished light absorption. Reflectivity showcased a decreasing trend with rising temperature, implying reduced reflection of incident light. Similarly, optical conductivity displayed a decline with temperature elevation, indicating weakened electric current conduction in response to incident light. The optical characteristics of Cubic $CH_3NH_3SnI_3$, characterized by transparency, stable refractive index, and relatively high reflectivity, position it as a promising candidate for solar cell applications. Additionally, its low optical conductivity, indicative of semiconductor behaviour, renders it suitable for efficient charge separation and transport within photovoltaic devices.

**Conflict of interest**
The authors declare no conflict of interest

**Declaration of Generative AI and AI Assisted Technologies in the Writing Process**
During the preparation of this work, the author(s) utilized OpenAI's GPT-3 language model to enhance the writing process. Subsequently, the author(s) thoroughly reviewed and edited the content as necessary, assuming full responsibility for the publication's content.